# Optimized multi-axis spiral projection MR fingerprinting with subspace reconstruction for rapid whole-brain high-isotropic-resolution quantitative imaging

Xiaozhi Cao[1,2], Congyu Liao[1,2,*], Siddharth Srinivasan Iyer[3,4], Zhixing Wang[5], Zihan Zhou[6], Erpeng Dai[1], Gilad Liberman[3], Zijing Dong[3,4], Ting Gong[6], Hongjian He[6], Jianhui Zhong[6,7], Berkin Bilgic[3,8,9] and Kawin Setsompop[1,2]

[1] Department of Radiology, Stanford University, Stanford, CA, USA

[2] Department of Electrical Engineering, Stanford University, Stanford, CA, USA

[3] Athinoula A. Martinos Center for Biomedical Imaging, Massachusetts General Hospital, Charlestown, MA, USA

[4] Department of Electrical Engineering and Computer Science, Massachusetts Institute of Technology, Cambridge, MA, USA

[5] Department of Biomedical Engineering, University of Virginia, Charlottesville, VA, USA

[6] Center for Brain Imaging Science and Technology, College of Biomedical Engineering and Instrumental Science, Zhejiang University, Hangzhou, Zhejiang, China

[7] Department of Imaging Sciences, University of Rochester, Rochester, NY, USA

[8] Department of Radiology, Harvard Medical School, Cambridge, MA, USA

[9] Harvard-MIT Division of Health Sciences and Technology, Massachusetts Institute of Technology, Cambridge, MA, USA

* Correspondence: Congyu Liao, PhD, cyliao@stanford.edu, Room 301, 350 Jane Stanford way, Stanford, CA, 94305 USA.

## Abstract

**Purpose**: To improve image quality and accelerate the acquisition of 3D MRF.

**Methods**: Building on the multi-axis spiral-projection MRF technique, a subspace reconstruction with locally low rank (LLR) constraint and a modified spiral-projection spatiotemporal encoding scheme termed tiny-golden-angle-shuffling (TGAS) were implemented for rapid whole-brain high-resolution quantitative mapping. The LLR regularization parameter and the number of subspace bases were tuned using retrospective in-vivo data and simulated examinations, respectively. $B_0$ inhomogeneity correction using multi-frequency interpolation was incorporated into the subspace reconstruction to further improve the image quality by mitigating blurring caused by off-resonance effect.

**Results**: The proposed MRF acquisition and reconstruction framework can produce provide high quality 1-mm isotropic whole-brain quantitative maps in a total acquisition time of 1 minute 55 seconds, with higher-quality results than ones obtained from the previous approach in 6 minutes. The comparison of quantitative results indicates that neither the subspace reconstruction nor the TGAS trajectory induce bias for $T_1$ and $T_2$ mapping. High quality whole-brain MRF data were also obtained at 0.66-mm isotropic resolution in 4 minutes using the proposed technique, where the increased resolution was shown to improve visualization of subtle brain structures.

**Conclusion**: The proposed TGAS-SPI-MRF with optimized spiral-projection trajectory and subspace reconstruction can enable high-resolution quantitative mapping with faster acquisition speed.

# Introduction

Magnetic resonance fingerprinting (MRF) (1) is an efficient quantitative imaging technique for estimating multiple tissue parameters simultaneously. The pulse sequence of MRF is usually designed to boost temporal incoherence by varying the acquisition parameters (e.g. flip angle (FA) and repetition time (TR)) and adding a preparatory module (e.g. inversion-preparation pulse and $T_2$-preparation pulse (2)). Therefore, voxels with specific tissue parameters, such as $T_1$ and $T_2$, can exhibit a unique temporal response to the MRF sequence, which is analogously termed as a "fingerprint". By matching the acquired "fingerprint" with a pre-calculated MRF dictionary, the tissue parameters could be retrieved from the best-matched entry, and thus the quantitative maps could be obtained by repeating this process voxel by voxel. Studies have demonstrated the repeatability and reproducibility of MRF in both quantitative relaxometry and morphometry (3–6). Initially demonstrated for quantitative neuroimaging, MRF also has been deployed and tailored for the imaging of other body parts, such as the heart (2,7), abdomen (8,9) and ankle (10). Since MRF enables rapid quantitative mapping for multiple tissue parameters, it has attracted significant interests as a diagnostic tool in diseases in which the pathological changes are sensitive to related quantitative parameters (11–17).

To further accelerate the acquisition and improve the image quality of MRF, a number of methods have been incorporated into the MRF framework from different aspects. One main direction among these is focused on the image reconstruction algorithms, including but not limited to the multi-scale reconstruction (18), the sliding-window method (19), low-rank based reconstruction techniques (20,21), spatiotemporal subspace modeling based reconstruction (20,22,23) and deep learning approaches (24,25). There have also been studies focused on correcting the reconstructed images influenced by system imperfections, such as imperfect slice-selection profile and $B_1^+$ inhomogeneity which could affect the accuracy of MRF quantitative mapping (26), and $B_0$ inhomogeneity which could induce blurring (27,28).

On the other hand, simultaneous multi-slice (SMS) (29,30) and 3D acquisition techniques (31–33) have been developed for MRF to achieve higher signal-to-noise (SNR) efficiency compared to 2D acquisition. With higher SNR efficiency, the 3D acquisition could achieve quantitative mapping with higher resolution, while at the same time the use of 3D encoding enables higher undersampling rate to obtain whole-brain mapping faster. For example, a method employing stack-of-spiral encoding scheme with interleaved undersampling in the partition-encoding dimension could achieve whole-brain MRF at 1.2×1.2×3 mm$^3$ resolution in 4.6 minutes (32). By augmenting this acquisition with a combined parallel imaging and deep learning reconstruction, 1-mm isotropic resolution whole-brain data can be acquired in 7 minutes (25). Another study has proposed to implement hybrid sliding-window and GRAPPA reconstruction on uniform undersampled stack-of-spiral MRF acquisition, which also achieves 1-mm whole-brain MRF in a similar time-frame (31). Rapid quantitative imaging is an active area of research and a number of other promising acquisition strategies have also been proposed for high-resolution whole-brain quantitative mapping, such as the multi-tasking technique (34) for 1×1×3.5 mm$^3$ resolution $T_1$/$T_2$/$T_1\rho$ mapping in 9 minutes and Echo Planer Time-resolved Imaging (EPTI) (35) for 1-mm isotropic resolution $T_1$/$T_2$/$T_2^*$ mapping in 3 minutes.

Our previous work proposed a novel multi-axis spiral-projection-imaging (SPI) acquisition scheme with sliding-window reconstruction to achieve 1-mm isotropic whole-brain quantitative mapping in 5 minutes, which was also demonstrated to be more robust to rigid head motions when compared to stack-of-spiral based acquisition schemes (33).

Building on the previous work, this study incorporates a spatiotemporal subspace modeling reconstruction with locally low rank (LLR) constraint and an optimized spatiotemporal sampling scheme into multi-axis SPI MRF. Compared to conventional sliding-window reconstruction, the subspace reconstruction is shown to provide markedly improved quantitative maps with lower artifacts and higher SNR. An optimized SPI encoding scheme is designed to increase spatiotemporal incoherence and validated to be more robust to artifacts, particularly at high acceleration rates. The proposed method enables high-quality whole-brain $T_1$, $T_2$, and proton density mapping at 1-mm isotropic resolution in less than 2 minutes. A 0.66-mm isotropic resolution scan can also be obtained in about 4 minutes, demonstrating the performance of the proposed method for submillimeter quantitative imaging. In addition, the multi-frequency interpolation (MFI) method (28,36) is also incorporated into the subspace reconstruction to mitigate blurring especially in large $B_0$ inhomogeneity regions to improve the image quality.

This work is an extension of our earlier work, which was reported in an abstract at the International Society of Magnetic Resonance in Medicine (ISMRM) 2021 (37).

## Method

### Sequence design

Figure 1A shows the pulse sequence diagram of the proposed method based on our previous work in (33). A conventional FISP-based MRF sequence (38) is employed, where 500 TRs with varying FAs (range from 10° to 90°, using non-selective Fermi pulses) are acquired after being prepared by an adiabatic inversion pulse with an inversion time of 20 ms. Each of such acquisition lasting 500 TRs is termed as an "acquisition group". When an acquisition group is completed, a resting time of 1.2 s is applied to allow for signal recovery before the next acquisition group. Different acquisition groups are used to performed different spiral encodings across multiple rotation axes to fill the 3D kspace. The TR and TE are fixed at 12 ms and 0.7 ms, respectively. Total 48 acquisition groups were used to achieve stable high-quality 1-mm whole-brain quantitative mapping in 5 minutes and 46 seconds.

In this work, when the number of acquisition groups equals to 3 times (for the purpose of multi-axis rotation scheme, where through-plane rotation was implemented on three orthogonal axes, $k_x$, $k_y$ and $k_z$) of center k-space undersampling factor of spiral, this acquisition was referred as an "$R=1$" acquisition for convenience and readability. But it does not mean that the data is literally fully sampled. For example, based on Nyquist sampling theory, the 3D undersampling factor of one single spiral is: in-plane acceleration factor × image matrix $/2\times\pi = 16\times220/2\times\pi = 552.6$. Even with 48 acquisition groups, the undersampling factor of each time point is 552.6/48=115.1, which would be even higher for bigger matrix size of higher resolution imaging. Therefore, sufficient k-space encoding was believed to be achieved by the massive TRs, for example 500 TRs in this work.

encoding was believed to be achieved by the massive number of TRs, for example 500 TRs in this work. Based on $R=1$, faster acquisitions such as $R=3$ and $R=6$ were achieved by undersampling acquisition groups.

In each TR, a variable density spiral (VDS) trajectory (39,40) with 16-fold in-plane undersampling rate at the center of k-space and a linearly increasing undersampling rate up to 32-fold at the edge of k-space is employed to achieve an encoding at 1-mm in-plane resolution and a FOV of 220×220 mm$^2$ with a readout duration of 6.7 ms. The high in-plane acceleration is used to alleviate the potential blurring caused by off-resonance phase accumulated during the readout duration. The slew rate and maximum gradient amplitude used for this spiral trajectory is 100 T/m/s and 40 mT/m, respectively, based on the consideration of compatibility for common clinic scanner.

**Faster acquisition with higher slew rate**

To achieve a faster acquisition with spiral trajectory, a higher slew rate of 150 T/m/s was used on the GE 3T Premier scanner (with maximum gradient amplitude of 80 mT/m and slew rate of 200 T/m/s) at the peripheral nerve stimulation (PNS) threshold. Similar readout duration (7.0ms here compared to 6.7ms as mentioned above), same 1-mm in-plane resolution and 220×220 mm$^2$ could be achieved using only 12-fold in-plane undersampling rate for center k-space and 24-fold for edge k-space. Therefore, only 36 (12×3 for multi-axis rotation scheme) acquisition groups were needed, resulting in a total acquisition time of 4 minutes and 19 seconds at $R=1$, which is 25% faster than the counterpart using slew rate of 100 T/m/s. Therefore, the acquisition time at $R=3$ using higher slew rate just needs 1 minute and 26 seconds, identical to $R=4$ using 100 T/m/s.

To assess the performance of the proposed method for submillimeter quantitative imaging, 0.66-mm isotropic resolution SPI-MRF was performed using a 6.8-ms VDS trajectory with 24-fold in-plane undersampling rate at the center of k-space and 48-fold at the edge of k-space using slew rate of 150 T/m/s. The number of acquisition groups was also increased to 72, resulting in a total acquisition time of 8 minutes 38 seconds at $R=1$.

The spiral readout durations were kept in a similar range (around 6.7~7ms) across different acquisitions to alleviate the effect of $B_0$ inhomogeneity and avoid substantial changes in sequence parameters such as TR.

**Optimized spatiotemporal k/t trajectory**

Figure 1B shows the spiral-projection k-space sampling encoding across the first 3 TRs and 16 acquisition groups for i) the previously used tiny-golden-angle (TGA) scheme (33) and ii) the proposed tiny-golden-angle-shuffling (TGAS) scheme. The TGA scheme performs in-plane rotations (i.e., spiral interleaving) in the group dimension and through-plane rotations in the TR dimension. On the other hand, to increase spatiotemporal incoherence, the TGAS scheme intermixes through-plane rotation along both the TR and the group dimensions. Figure 1C shows the color-coded plots of all the spirals that are played out during the first 3 TRs for the TGA and the TGAS schemes.

As shown in the left panel of Figure 1C, for the first 16 acquisition groups ($G_{1-16}$) in the TGA scheme, in-plane rotations (around the z-axis) are performed along the acquisition group dimension to obtain 16 spiral interleaves and achieve full sampling at the center of k-space, to provide a disk-like coverage in 3D k-space. Through-plane rotations (around the x-axis) are then performed along the TR dimension using the tiny golden angle step (41) of 23.63°. Thus the disk-like k-space coverage is gradually rotated to cover the entire 3D k-space along with TRs. With the TGA scheme, the in-plane rotation angle $\theta_z$ and the through-plane rotation angle $\Theta_x$ could be described as: $\Theta_x(j) = 23.63° \times j, \theta_z(i) = 22.5° \times i$, where $j$ is the TR index and $i$ is the index of the acquisition group. Here, the angle of 22.5° is determined by the in-plane undersampling rate at the center of k-space of a single spiral interleaf, namely 360°/16 here for 1-mm isotropic resolution, which would be updated to 360°/24 for the 0.66-mm version. To achieve multi-axis rotations, groups $G_{17-32}$ and $G_{33-48}$ are designed to rotate around x- and y-axis for in-plane rotation and y- and z-axis for through-plane rotation, respectively (shown on the left-side of Figure 1D).

In this work, we propose the TGAS spatiotemporal encoding scheme which shuffles the rotation operations via: $\Theta_x(i,j) = 23.63° \times (i+j), \theta_z(i) = 22.5° \times i$, as illustrated in the right panel of Figure 1B. Here, the in-plane and through-plane rotations are intermixed in both the TR and the acquisition group dimensions. This increases the spatiotemporal incoherency, which should be aid the subspace reconstruction. The advantage of the TGAS over TGA scheme was validated by both simulation and in-vivo experiments in this work. Similar to TGA, TGAS also employs the multi-axis rotations as shown on the right-side of Figure 1D.

**Subspace reconstruction**

Figure 2A shows the process of the MRF subspace reconstruction. The MRF dictionary was pre-calculated using the extended phase graph (EPG) method (42). The resulting three-dimensional dictionary ($T_1$ entries, $T_2$ entries and TRs, respectively) was reshaped to two-dimensions (entries including all $T_1$&$T_2$ and TR) and a singular value decomposition (SVD) was then applied on the reshaped dictionary. The first five temporal (TR) principal components were extracted as subspace bases (22,43–47), $\Phi_{1-5}$. The coefficient maps ($c_{1-5}$) of the bases could then be solved by:

$$\min_{c}\|PFS\Phi c - y\|_2^2 + \lambda R_r(c) \qquad [1]$$

where $P$ is the undersampling pattern, $F$ is the non-uniform Fourier (NUFFT) transform, $S$ are the coil sensitivity maps, $y$ are the acquired raw data, $\lambda$ is the regularization parameter for locally low rank regularization (LLR) (44,48) $R_r(c)$. The coil sensitivity maps $S$ were estimated with the ESPIRiT method (49) using the central k-space data combined from all acquired TRs and groups. The reconstruction was implemented using the BART toolbox (50).

In order to optimize the selection of $\lambda$ for LLR regularization used in the subspace reconstruction, a reference $T_1$ and $T_2$ maps at 1-mm isotropic resolution were created from an oversampled TGAS-SPI-MRF dataset (termed as $R$=0.5, which corresponds to a 11 minutes and 31 seconds acquisition at 1-mm resolution, with 96 non-overlapping acquisition groups) using subspace reconstruction without regularization. Retrospective undersampling of this $R$=0.5 dataset was then performed along the acquisition group dimension to generate $R$=3 and $R$=6 datasets, that were then reconstructed with the subspace approach using $\lambda$ ranging from $10^{-6}$ to $10^{-4}$. The difference maps and root-mean-square-error (RMSE) of the resulting $T_1$ and $T_2$ maps relative to the reference maps were calculated (shown in the Supporting Figure S1 and S2). Based on the consideration of SNR and image quality, $\lambda$ $3\times10^{-5}$ and $5\times10^{-5}$ were selected for $R$=3 and $R$=6, respectively.

In order to determine the number of subspace bases used in the reconstruction, a simulation experiment was performed to test the reconstruction performance with subspace bases ranging from 3 to 7; the results of which are shown in the Supporting Figure S3. The simulation data were generated using $T_1$, $T_2$ and PD maps obtained from the above $R$=0.5 TGAS-SPI-MRF acquisition and these quantitative maps were also used as reference image for calculating the RMSE. Since reconstructing with more subspace bases requires more reconstruction time and more random-access memory (RAM), the number of subspace bases was set to 5 to achieve sufficient accuracy under the available hardware and reconstruction time constraints.

To accelerate the reconstruction and reduce memory requirement, SVD coil compression (51) was employed for all reconstructions to compress the effective channel count to 12 (for both 48-channel and 64-channel coils used in this work).

To provide a baseline comparison for the subspace reconstruction, the conventional sliding-window inverse NUFFT (iNUFFT) reconstruction (33) was also implemented, comprising the following steps: i) combine data acquired from same TR of all acquisition groups together directly; ii) apply sliding-window approach along TR dimension with a window width of 50 to get sphere-like k-space coverage for each sliding-windowed time point.; iii) apply 3D iNUFFT operator (52) to the sliding-windowed data to reconstruct mixed-contrast and approximately fully-sampled images.

**Simulation**

In this work, two simulation experiments were performed: one for determining the number of subspace bases, which has been described above, and the other for the quantitative performance analysis between the TGA and TGAS acquisition schemes. The $T_1$, $T_2$ and PD maps created from a $R$=0.5 TGAS-SPI-MRF dataset described above were used as reference and for generating the synthetic MRF time-series images along TR dimension. To generate simulation data for TGA and TGAS acquisitions, these synthetic images were added with white noise (the amplitude of noise is 25% of the average amplitude of synthetic images) in image domain and passed through the forward model comprising coil sensitivities and k-space trajectory sampling. The error maps and RMSE were calculated by using these simulated data.

**Template match**

The subspace bases compression is also applied to the MRF dictionary, which substantially decreased the matrix size of the dictionary from 160×176×500 to 160×176×5, where 160 is the number of $T_1$ entries (corresponding to [20:20:3000, 3200:200:5000] ms), 176 is for $T_2$ (corresponding to [10:2:200, 220:20:1000, 1050:50:2000, 2100:100:4000] ms), 500 is the number

of TRs within an acquisition group and 5 is the number of subspace bases. Therefore, the reconstructed coefficient maps could be used directly for template matching with the compressed dictionary, which significantly improves the computation speed of the template matching(53).

**B₀ correction**

To further improve the image quality, the multi-frequency interpolation (MFI) technique(28,36) was incorporated into the subspace reconstruction, with Equation [1] updated as:

$$\min_{c_m} \lVert PFS\Phi c_m - e^{-it\omega_m}y \rVert_2^2 + \lambda R_r(c_m) \quad [2]$$

where $e^{-it\omega_m}$ is the conjugate phase demodulation term with time $t$ accumulated during the readout duration at a specific frequency $\omega_m$. In this work, 5 demodulation frequencies $\omega_{1\sim5}$ = [-200:100:200] Hz were used to cover the typical B₀ inhomogeneity range in human neuroimaging on common clinical 3T scanners. Therefore, the coefficient maps with MFI correction $c_{\text{MFI}}$ could be obtained by:

$$c_{\text{MFI}}(r) = \sum_{m=1}^{M} W_m(r) c_m(r) \quad [3]$$

where $r$ is the voxel position, $M=5$ corresponds to $\omega_{1\sim5}$ and $W_m(r)$ is the weighting coefficient maps which is the linear interpolation factor of $\omega_{1\sim5}$ to match the measured B₀ inhomogeneity $\Delta B_0$ at $r$.

**B₁⁺ correction**

To improve the accuracy of MRF quantification and account for B₁⁺ inhomogeneity effects in the MRF data, signal evolution across a range of discretized B₁⁺ values ([0.50:0.05:1.50], i.e. ±50% B₁⁺ variations) were simulated using the EPG method, which enabled the creation of a set of MRF dictionaries with different B₁⁺ variations. The corresponding dictionary was then selected for use in each spatial location based on a pre-scanned B₁⁺ map (26).

**In-vivo validation**

To validate our proposed method, five healthy volunteers were scanned with the approval of Institutional Review Board. Studies were performed on two 3T MAGNETOM Prisma scanners (Siemens Healthcare, Erlangen, Germany) with a 32-channel head receiver coil and a 3T Premier MRI scanner (GE Healthcare, Madison, WI) with a 48-channel head receiver coil. The matchup between results shown in figures and scanners was reported in the Supporting Table S1.

An oversampled $R$=0.5 1-mm dataset with TGAS trajectory was acquired as a reference to optimize the selection of LLR regularization parameter $\lambda$ and the number of subspace bases used in the reconstruction. Both TGAS and TGA MRF data were also acquired at 1-mm resolution at $R$=1. Retrospective undersampling (along acquisition group dimension) experiments at acceleration factors of $R$=3 (1 m 55 s) and $R$=6 (58 s) were performed to validate the performance of the proposed method to provide a faster acquisition. In addition, 1-mm and 0.66-mm TGAS-SPI-MRF using higher slew rate (150 T/m/s) were also acquired.

To incorporate $B_0$ correction into the subspace reconstruction, an additional $B_0$ field map was acquired using a multi-echo gradient-echo scan with two different TEs (2.6 ms and 4.1 ms) within 1 minute. The FOV was 220×220×220 mm$^3$ to match the MRF acquisition but with a lower resolution of 3.4×3.4×4 mm$^3$ on the grounds that the $B_0$ field inhomogeneity is relatively smooth and higher SNR is preferred rather than higher resolution.

To correct the $B_1^+$ inhomogeneity in MRF data, an FOV-matched $B_1^+$ map was obtained by using a vendor-supplied $B_1^+$ mapping sequence (54). The FOV was 220×220×220 mm$^3$ and the scan time for this $B_1^+$ mapping at 3.4×3.4×5 mm$^3$ resolution was about 60 seconds.

To analyze the $T_1$ and $T_2$ values by using different methods or acquired on different MRI systems, synthetic $T_1$-weighted images generated by using the measured $T_1$ and PD maps from MRF were used as input for FreeSurfer toolbox (55) for brain segmentation. The average $T_1$ and $T_2$ values of GM and WM on both bilateral hemispheres were then calculated based on this segmentation.

Computations were performed on a Linux (Ubuntu 20.04) server (with 32 Core i7 Intel Xeon 2.8 GHz CPUs, an Nvidia 2080Ti GPU and 512GB RAM) using MATLAB R2015b (The MathWorks, Inc., Natick, MA).

## Results

Figure 2B shows comparisons between the sliding-window iNUFFT and the subspace reconstructions for 1-mm MRF data. At $R=3$ with TGA acquisition, the $T_1$ and $T_2$ maps from the sliding-window iNUFFT reconstruction show significant artifacts, while the $T_1$ and $T_2$ maps from the subspace reconstruction are of good quality, with arguably better quality than ones from sliding-window iNUFFT reconstruction at $R=1$. However, as indicated by the yellow arrow, there are some residual artifacts in the $T_2$ map from the TGA acquisition at $R=3$ with subspace reconstruction. For an equivalent TGAS acquisition with subspace reconstruction at $R=3$, this artifact absent.

Figure 3 further compares the performance of 1-mm TGA and TGAS acquisitions, both with subspace reconstruction, at various acceleration factors. The results from TGA show slight artifacts at $R=3$ (indicated by yellow arrows with zoom-in regions) and strong artifacts at $R=6$ (red arrows). On the other hand, the results from TGAS remain artifact-free and retain high image quality, indicating the superiority of the TGAS acquisition scheme. The top section of Table 1 reports the average $T_1$ and $T_2$ values in the white-matter (WM) and gray-matter (GM) in subject 1, obtained using three different combinations of acquisition and reconstruction schemes. The reported values are consistent across methods, in particular with the TGA trajectory and sliding-window iNUFFT reconstruction having previously been validated to agree well with gold-standard quantitative imaging techniques (33). This indicates that the $T_1$ or $T_2$ maps are consistent when subspace reconstruction and the TGAS trajectory are employed.

Figure 4 shows the simulation results from subspace reconstructions of synthesized TGA and TGAS data at 1-mm isotropic resolution, at $R=3$ and $R=6$. The $T_1$ and $T_2$ reference maps obtained from $R=0.5$ TGA data are also shown for comparison. Similar to the in-vivo results, results from TGA show slight artifacts at $R=3$ and strong artifacts at $R=6$ while results from TGAS remain artifact-free. In comparing with the reference, the results from TGAS shows lower RMSE than those from TGA, for both $T_1$ and $T_2$ maps at $R=3$ and $R=6$.

Figure 5 shows the $T_1$ and $T_2$ maps of 1-mm and 0.66-mm TGAS-SPI-MRF using 100 T/m/s slew rate and 150 T/m/s slew rate. To compare the results of 1-mm resolution with identical acquisition time (1 minute 26 seconds), $R=4$ and $R=3$ were used for low and high slew rate, respectively. Indicated by yellow arrows, using higher slew rate and gradient amplitude could help improve quantitative mapping quality of MRF within same acquisition time. Based on SNR consideration, $R=2$ (4 minutes 19 seconds) was selected for the acquisition with 0.66-mm isotropic resolution.

The increased resolution of the 0.66-mm acquisition can help better visualize subtle brain structures indicated by red arrows (from left to right: claustrum, small sulci, caudate nucleus shown on the $T_1$ maps; optic radiation, medial occipitotemporal gyrus and cerebellum shown on the $T_2$ maps, respectively), where the zoom-in figures highlight details in several specific regions.

Figure 6 compares the results with and without MFI $B_0$ correction for 1-mm TGAS-SPI-MRF at $R=3$. It can be seen that in regions with strong $B_0$ inhomogeneity (for example, $\Delta B_0$ is around 150 Hz indicated by the red arrow in the top row), the $T_1$ and $T_2$ maps show substantial distortion. In regions with $\Delta B_0$ around 80Hz (shown in the middle row), recognized blurring could be found. And in regions with $\Delta B_0$ around 30Hz (shown in the bottom row) or less, neither distortion or aliasing could be recognized by comparing the $T_1$ and $T_2$ maps with and without $B_0$ correction. An MPRAGE acquisition (the 6th column of Figure 6) was used as reference to demonstrate the effectiveness of $B_0$ correction indicated by the zoom-in comparison (the 7th column).

Supporting Figure S4 shows the results of $B_1^+$ corrections in MRF reconstruction. With the $B_1^+$ correction, the estimated $T_2$ maps in regions indicated by red arrows are more uniform compared to the standard MRF reconstruction without $B_1^+$ corrections.

Figure 7 shows the quantitative maps obtained from a single subject scanned at two different sites months apart, using 3T scanners from two different vendors (acquired at $R=3$ for 1-mm TGAS-SPI-MRF). In particular, the subject was scanned on both a Siemens 3T Prisma scanner (Martinos Center, Massachusetts General Hospital, Boston, MA, USA and a GE 3T Premier scanner (Lucas Center, Stanford University, Stanford, CA, USA), where the results from these scans are shown in the first and second rows, respectively. The difference images obtained after co-registering (using FSL MCFLIRT (56)) these parameter maps from the two scans are shown in the bottom row with 10x scaling. Minimal differences in the GM and WM regions are observed across these scans. The average quantitative values of the bilateral WM and GM are shown in the Table 1 under Subject 2.

## Discussion

In this study, a subspace reconstruction with optimized spatiotemporal trajectory termed TGAS-SPI-MRF was proposed to improve the image quality and accelerate the acquisition speed of SPI-based 3D-MRF. By projecting the data along the temporal dimension on to a low-dimension subspace and applying LLR regularization, the reconstruction conditioning as well as the SNR are significantly improved, allowing for higher accelerations. An optimized shuffled 3D spiral-projection trajectory with improved spatiotemporal incoherency further improves the quantitative maps at high acceleration factors of $R$=3 and $R$=6, corresponding to acquisition times of about 2 and 1 minutes for 1-mm isotropic resolution, respectively. The results were validated on both the GE Premier and the Siemens Prisma 3T scanners, indicating the robustness and reproducibility of the proposed method. $B_0$ correction using MFI was also incorporated into the subspace reconstruction to mitigate blurring in regions with strong $B_0$ inhomogeneity. Moreover, submillimeter quantitative mapping at 0.66-mm isotropic resolution was also achieved in 4.3 minutes and was shown to better delineate subtle brain structures. The ability of the proposed technique to provide rapid high resolution quantitative neuroimaging should help facilitate the adoption of quantitative imaging in broad neuroscientific and clinical applications.

By comparing the results from sliding-window iNUFFT and subspace reconstructions as well as TGA and TGAS trajectories, both the quantitative maps and values demonstrated that neither the subspace reconstruction nor the TGAS trajectory induce bias for $T_1$ and $T_2$ mappings. Since the accuracy of 3D MRF using TGA trajectory and sliding-window iNUFFT reconstruction has been validated in our previous work (33), we did not perform additional phantom examination in this work.

The subspace reconstruction and LLR regularization improve the reconstruction conditioning as well as SNR. To balance the tradeoff between SNR and image sharpness, the regularization parameter ($\lambda$) of $3 \times 10^{-5}$ and $5 \times 10^{-5}$ were selected for $R$=3 and $R$=6, respectively. Moreover, five subspace bases were utilized in this work to balance between achievable image quality and computation hardware limitation. Both of these parameters ($\lambda$ and the number of subspace bases) were used consistently in all of the reconstructions shown in this study to achieve stable image quality. This demonstrates that properly tuned selection of these parameters can enable robust results with limited spatial blurring.

One limitation of the proposed method is the computation hardware requirement and the long reconstruction time. Although a coil compression method was employed, the reconstruction still took 3-5 hours on a Linux Server and required a minimum of 200GB of RAM. To mitigate this issue, future work will focus on the development of an efficient reconstruction algorithm with a small memory footprint that can leverage distributed GPU computing. A promising approach in this direction is the use of stochastic optimization that has recently been applied to a large spatiotemporal MRI reconstruction problem (57). The utilization of deep learning techniques in place of the subspace reconstruction could be another potential solution, which could also help denoised the image to allow for higher resolution imaging in a faster acquisition time (24).

Gradient hardware imperfections can cause image artifacts and impact achievable spatial resolution. Conventional golden-angle trajectory would rapidly change the eddy currents, which could induce signal fluctuation and impair image quality. It has been demonstrated that for SPI acquisition (41) the use of TGA can help minimize shot-to-shot differences caused by eddy currents, and this have been adopted in our previous work on TGA-SPI-MRF (33). In this work, the TGAS trajectory was proposed to increase the spatiotemporal incoherency of the data acquisition. Compared to TGA, which performs a through-plane rotation of 23.63° between 2 adjacent TRs, the TGAS performs this rotation as well as an extra in-plane rotation of 22.5°. This added rotation is still relatively small compared with conventional golden angle of 111.25°, which should help to limit the increase in shot-to-shot difference caused by eddy currents as supported by the high-quality quantitative mapping achieved with this approach. Additionally, future work will also explore the use of field probes (58) to achieve detailed characterization of eddy current and gradient trajectory imperfections. This information can then be incorporated into the reconstruction (59) to further improve the image quality and parameter quantitation. Such characterization should help open up the possibility to explore more varied spatiotemporal trajectory to further increase spatiotemporal incoherency and enable higher accelerations.

The imperfections from $B_0$ and $B_1^+$ inhomogeneities could also affect the performance of MRF mapping. $B_0$ correction based on the MFI method has been incorporated into the subspace reconstruction and shown to be effective at mitigating blurring caused by off-resonance effect. The imaging results show that even with highly segmented spirals, the image quality can still be degraded by severe $B_0$ inhomogeneity, as in the case when $\Delta B_0$ is higher than 150 Hz. Nonetheless, this can be corrected well using MFI, which will be important in applying TGAS-SPI-MRF to other body parts where $B_0$ inhomogeneity is significantly larger than in the brain. However, the approach requires an additional $B_0$ map acquisition which took ~ 1 minute in this work for whole brain imaging. Moreover, the results from this study and others (26,33) have demonstrated that

$B_1^+$ correction can improve the quantitative accuracy of MRF (26), especially for $T_2$ mapping (since the inversion pulse in an MRF sequence is typically adiabatic, the $T_1$ measurement is less affected than $T_2$ which relies more on the varying flip angles excitation train). This will also require additional acquisition time. To address this issue, the use of rapid $B_0$ and $B_1^+$ mapping will be explored where the PhysiCal is a promising approach which has been recently demonstrated to provide high quality whole brain mapping of coil sensitivity, $B_0$, and $B_1^+$ simultaneously in 11 seconds (60). Another direction is to incorporate the mapping of these additional parameters directly into the MRF sequence design and reconstruction, as per the MR field fingerprinting approach (61).

TGAS-SPI-MRF provides a basis for extremely efficient spatiotemporal sampling, not only for $T_1$ and $T_2$ mapping as demonstrated here, but can also be extended to other applications such as diffusion-relaxometry and microstructure mapping such as with myelin-water which we have started to explore (62). It could also be adapted for other time-series imaging applications such as perfusion imaging with subspace or model-based reconstruction. Given the short TE and motion robustness features of SPI-based trajectory (63), this method should be amendable to body imaging, especially for motion-sensitive parts, as well.

The application of TGAS-SPI-MRF at ultra high-field, such as at 7T, would enable a significant increase in SNR to push the achievable spatial resolution to a finer scale and provide robust quantitative assessment of tissue parameters across cortical depths and fine-scale substructures (64,65). Since the encoding speed of spiral trajectory is mainly limited by slew rate, novel high-performance gradient systems (66–68) with higher slew rate and PNS threshold could bring benefits to this method as well. The results in this work have shown that utilizing higher slew rate can enable faster k-space transversal to reduce the number of spiral interleaves required in the TGAS-SPI-MRF acquisition while keeping the spiral readout duration constant. This allows for faster acquisition and/or the ability to obtain better image quality within same acquisition time. Based on our gradient trajectory and PNS simulations, the acquisition time of TGAS-SPI-MRF at both 1-mm and 0.66-mm resolutions can be further reduced by 2-3 fold if the usable slew rate is increased from 100-150 T/m/s used in this work to 500-800 T/m/s, something achievable on emerging novel gradient systems. Therefore, the combined use of advanced gradient system and ultra high-field MRI for TGAS-SPI-MRF acquisition could offer an exciting possibility in rapid mesoscale quantitative imaging of the brain.

## Conclusion

A rapid, high-resolution, whole-brain MRF technique was developed using the optimized tiny-golden-angle-shuffling acquisition scheme and subspace reconstruction with locally low rank regularization. The proposed method can obtain high-quality whole-brain $T_1$, $T_2$ and proton density maps with 1-mm isotropic resolution in 2 minutes at $R=3$ with similar image quality as its $R=1$ counterpart, or even 1 minute at $R=6$ with slight compromise on image sharpness. The 0.66-mm submillimeter quantitative mapping has also been achieved with an acquisition time of 4 minutes and 19 seconds. MFI-based $B_0$ correction is also incorporated into the reconstruction to mitigate the blurring caused by off-resonance effect. In-vivo experiments on healthy volunteers have been validated across both GE and Siemens platforms. With its fast, accurate, and high-resolution acquisition and motion robustness, this method may be valuable for various clinical applications that require fast high-resolution quantitative $T_1$ and $T_2$ mapping and for subjects susceptible to motion.

## Acknowledgement

The authors would like to acknowledge Dr. Stefan Skare for his supporting on sequence programming based on KS Foundation platform (https://ksfoundationepic.org/).
This work was supported by:
NIH research grants: R01EB020613, R01MH116173, R01EB019437, R01EB028797, R01EB016695, U01EB025162, P41EB030006, U01EB026996, R03EB031175.

**Table**

|           |                | $T_1$   |         |         |         | $T_2$   |         |         |         |
|-----------|----------------|---------|---------|---------|---------|---------|---------|---------|---------|
|           | Unit: ms       | WM (L)  | WM (R)  | GM (L)  | GM (R)  | WM (L)  | WM (R)  | GM (L)  | GM (R)  |
| Subject 1 | TGA+iNUFFT     | 883.2   | 878.3   | 1362.2  | 1366.7  | 60.33   | 59.55   | 81.54   | 80.51   |
|           | TGA+SubRecon   | 874.3   | 865.9   | 1373.8  | 1362.1  | 61.28   | 62.62   | 82.09   | 82.28   |
|           | TGAS+SubRecon  | 857.5   | 845.4   | 1362.0  | 1358.8  | 61.92   | 60.45   | 81.83   | 80.01   |
| Subject 2 | TGAS(Siemens)  | 843.5   | 838.1   | 1323.7  | 1328.1  | 62.71   | 62.13   | 81.18   | 80.51   |
|           | TGAS(GE)       | 847.7   | 849.1   | 1340.6  | 1352.0  | 62.03   | 60.56   | 80.62   | 79.88   |

Table 1.

$T_1$ and $T_2$ values of WM and GM estimated by using different method (the top 3 rows) and across different scanners (the bottom 2 rows) with same MRF acquisition protocal.

# Figures

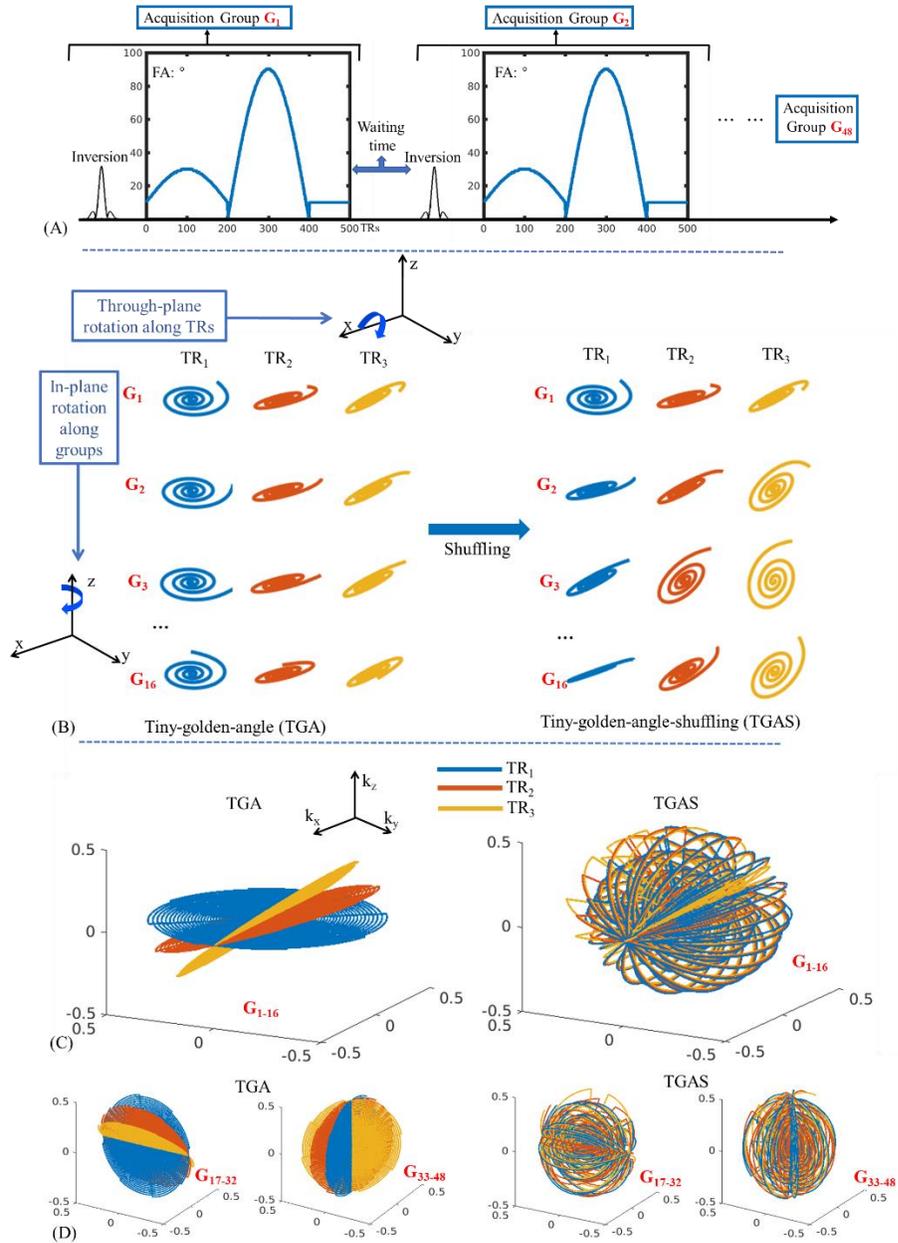

Figure 1.

(A) Sequence diagram.

(B) Spiral-projection spatiotemporal encoding with *left*) the original tiny-golden-angle (TGA) scheme and *right*) the proposed tiny-golden-angle-shuffling (TGAS) scheme.

(C) and (D) the k-space coverage of the first 3 TRs for acquisition groups $G_{1-16}$, $G_{17-32}$ and $G_{33-48}$, where through-plane rotation was implemented around x-, y- and z-axis respectively to achieve multi-axis rotation for better incoherence.

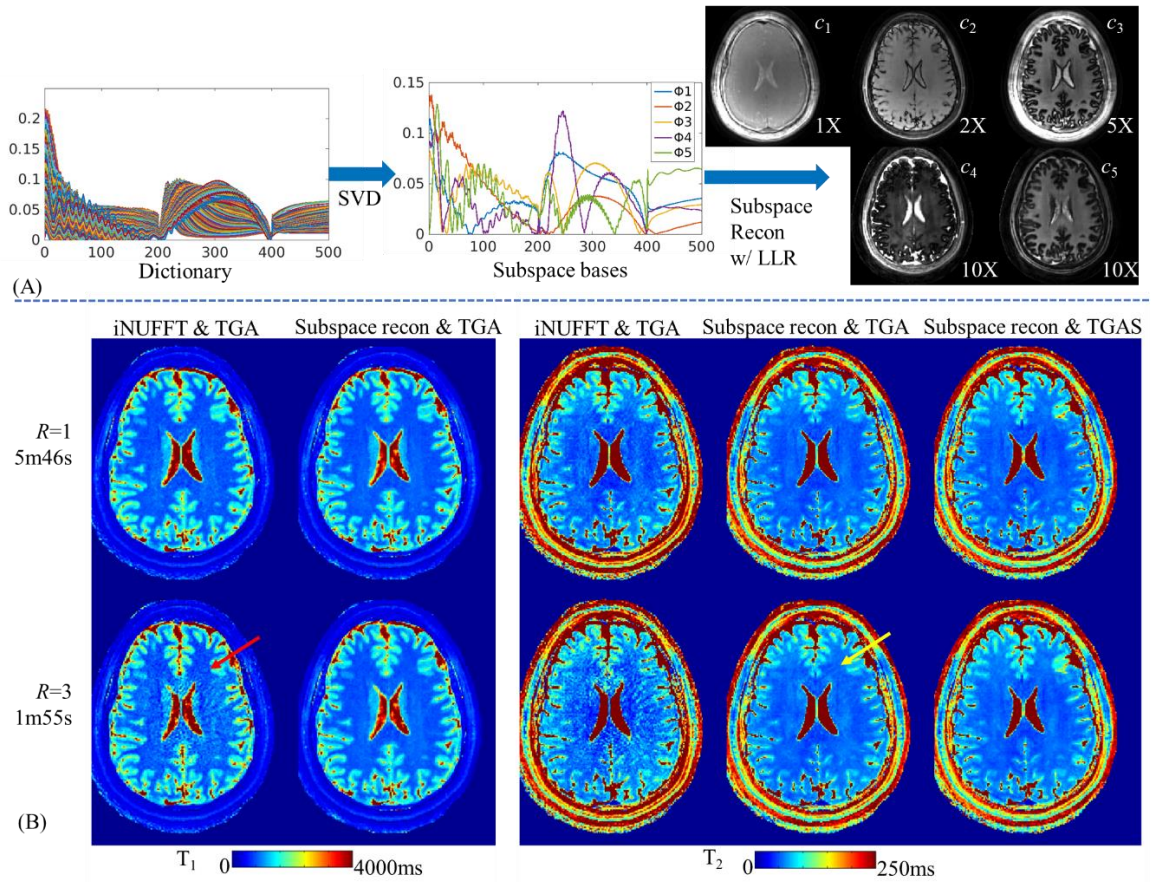

Figure 2.

(A) The Schematic workflow of subspace reconstruction. Five subspace bases were extracted from the MRF dictionary and used to reconstruct the coefficient maps (at 1×, 2×, 2×, 10×, and 10× scaling respectively for better visualization). The coefficient maps are then used to generate the MRF time-series images and dictionary template matching performed to obtain $T_1$, $T_2$, and PD maps.

(B) Comparison between sliding-window iNUFFT and subspace reconstructions, where $T_1$ & $T_2$ maps from 1-mm isotropic acquisitions at acceleration factors $R=1$ and $R=3$ are shown.

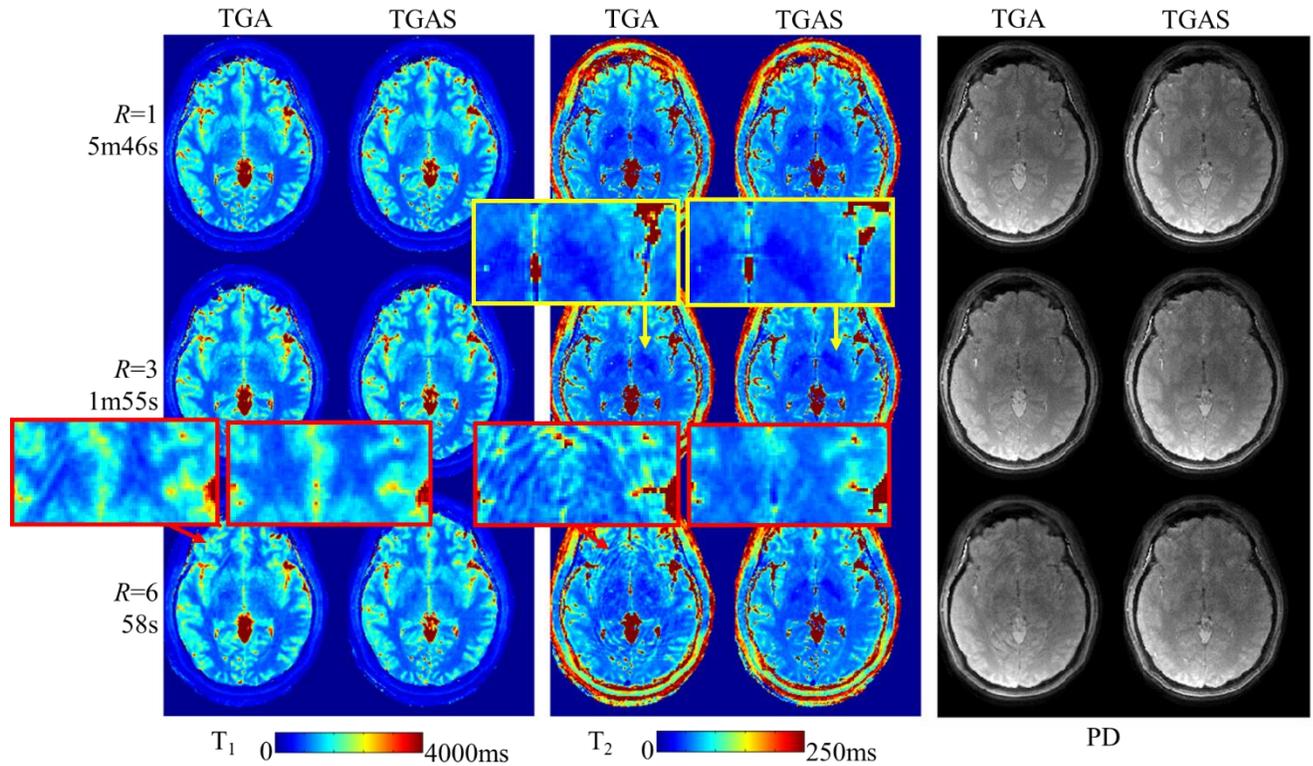

Figure 3.

In-vivo comparison between the conventional TGA and the proposed TGAS spatiotemporal encodings at $R=1$, 3, and 6, all with subspace reconstruction. The TGA version shows slight artifacts at $R=3$ (yellow arrow) and strong artifacts at $R=6$ (red arrow), while the proposed TGAS version maintains good image quality throughout.

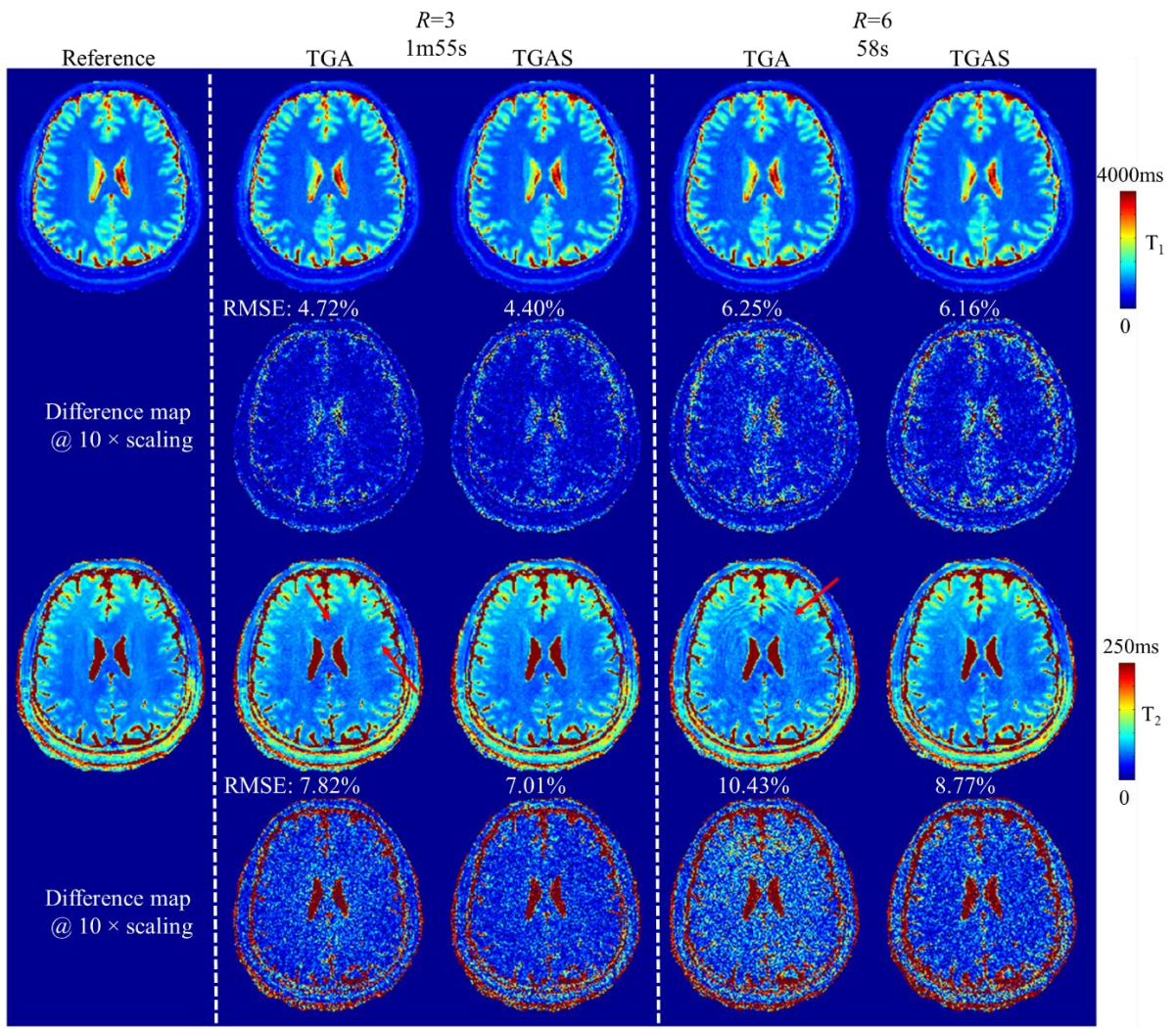

Figure 4.

Simulation comparison between the conventional TGA and the proposed TGAS spatiotemporal encodings at *R*=3 and *R*=6.

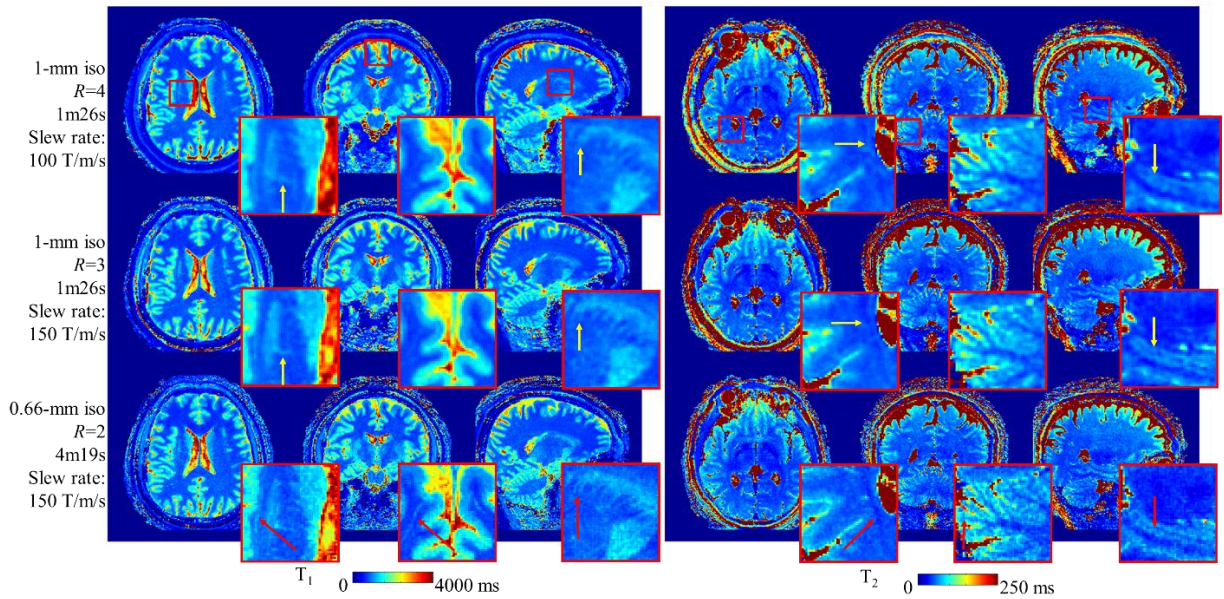

Figure 5.

$T_1$ & $T_2$ maps with zoom-in regions of the 1-mm and 0.66-mm isotropic resolution datasets. Two different slew rates (100 T/m/s for top row and 150 T/m/s for middle row, respectively) were acquired. *R*=4 was used for low slew rate acquisition to achieve an identical acquisition time with high slew rate version. Yellow arrows indicate the improvement by using higher slew rate for achieving better image quality within same acquisition time. The higher resolution in the 0.66-mm dataset can aid in better visualization of subtle brain structures as indicated by the red arrows.

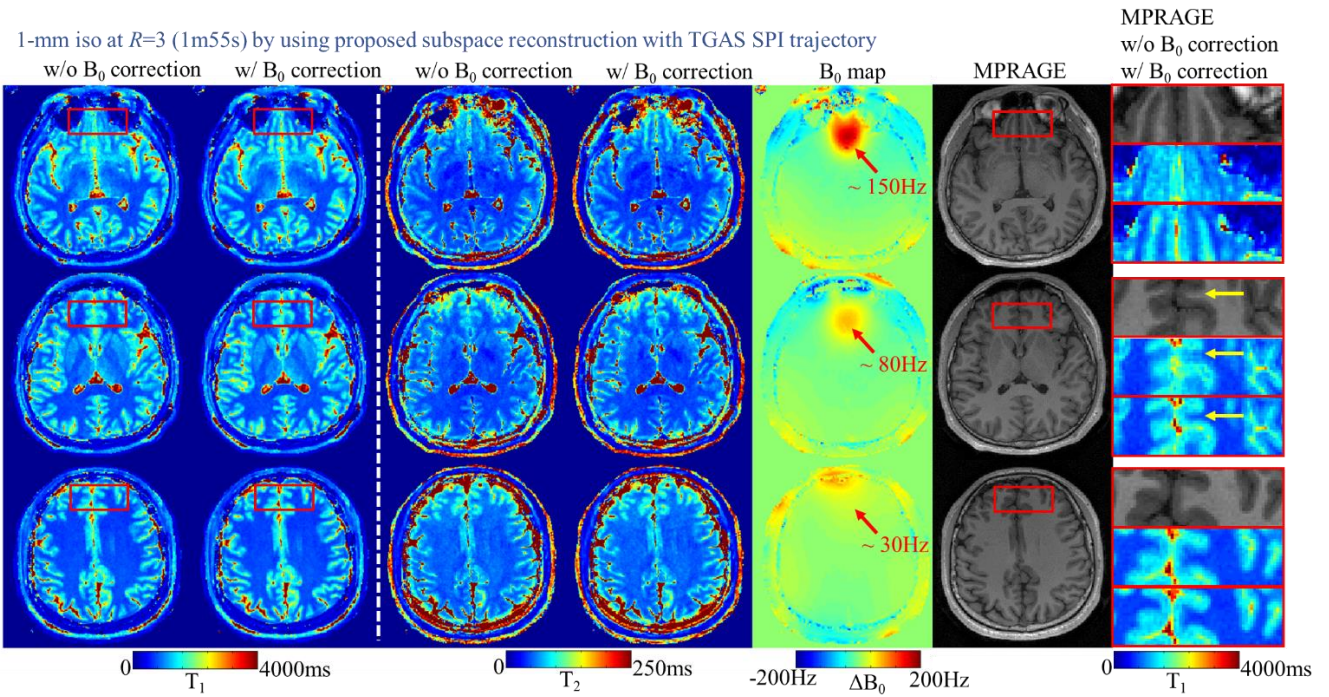

Figure 6.

$T_1$ & $T_2$ maps using the proposed TGAS-SPI-MRF with and without MFI $B_0$ correction. $B_0$ maps and MPRAGE images are also shown for better visualization of the $B_0$ correction effectiveness. The zoom-in images of MPRAGE, $T_1$ maps without $B_0$ correction and with $B_0$ correction are shown in the right-most from top to bottom, respectively.

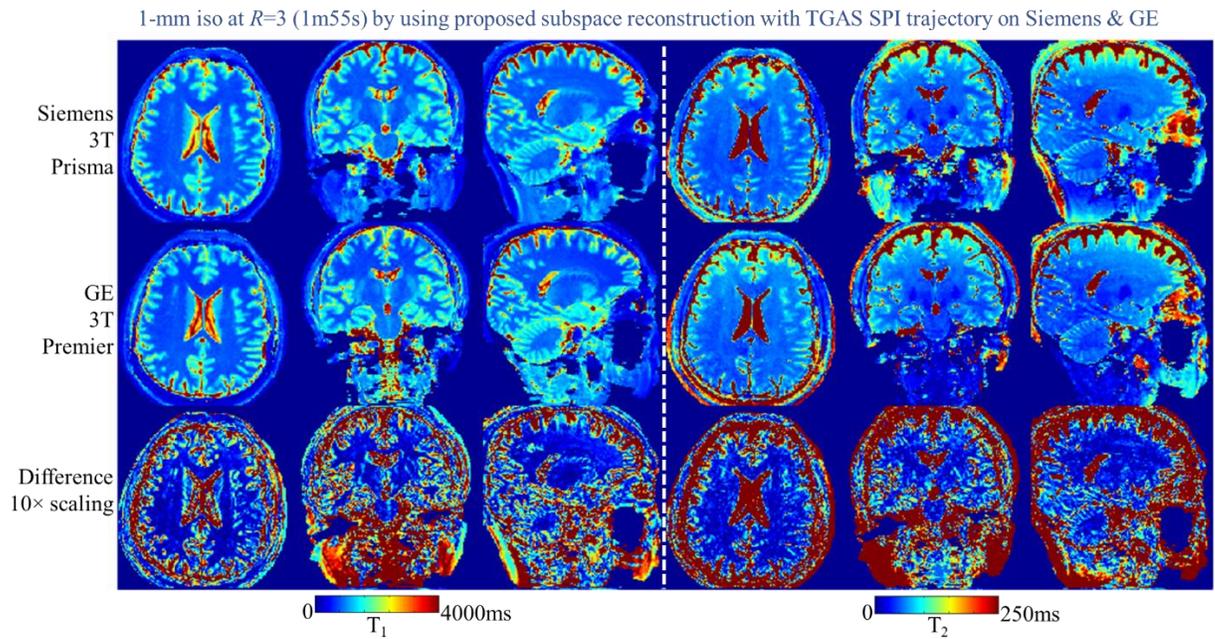

Figure 7.

T$_1$ and T$_2$ maps of one subject scanned on a Siemens 3T Prisma (top row) and a GE 3T Premier (middle row) using the proposed TGAS-SPI-MRF. The image was registered to a similar position by using FSL MCFLIRT to calculate the difference map (bottom row).

# Supporting information

|  | Siemens Prisma 1 | Siemens Prisma 2 | GE Premier |
|---|---|---|---|
| Figure 2/3/4 | √ |  |  |
| Figure 5 |  |  | √ |
| Figure 6 |  | √ |  |
| Figure 7 | √ |  | √ |
| Figure S1/S2/S3 | √ |  |  |
| Figure S4 |  |  | √ |

Supporting Table S1.

Matchup between results shown in figures and their corresponding scanners. The Siemens Prisma 1 is located at Martinos Center, Massachusetts General Hospital, Boston, MA, USA, Siemens Prisma 2 at Center for Brain Imaging Science and Technology (CBIST), Zhejiang University, Hangzhou, Zhejiang, China, and GE Premier scanner at Lucas Center, Stanford University, Stanford, CA, USA.

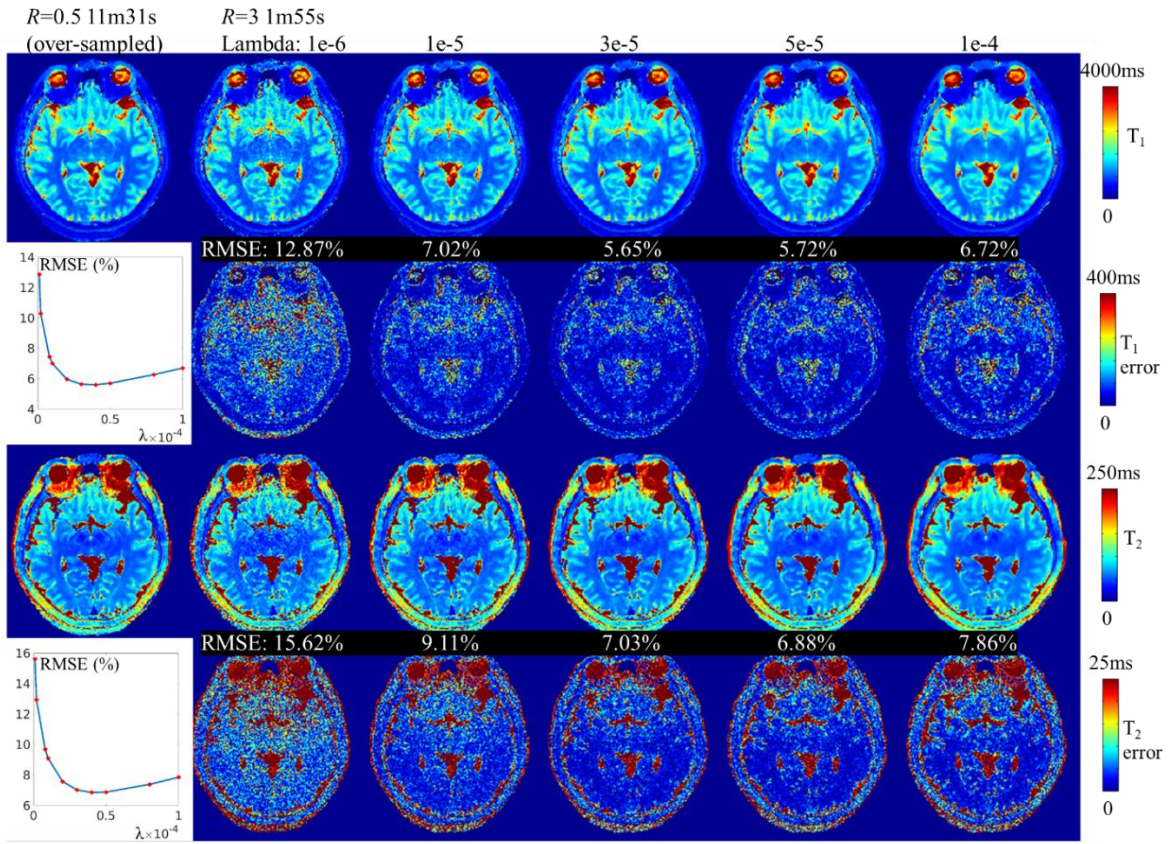

Supporting Figure S1.

$T_1$ and $T_2$ maps using the proposed TGAS-SPI-MRF at $R=3$ with LLR regularization parameter $\lambda$ from $10^{-6}$ to $10^{-4}$ using in-vivo data.

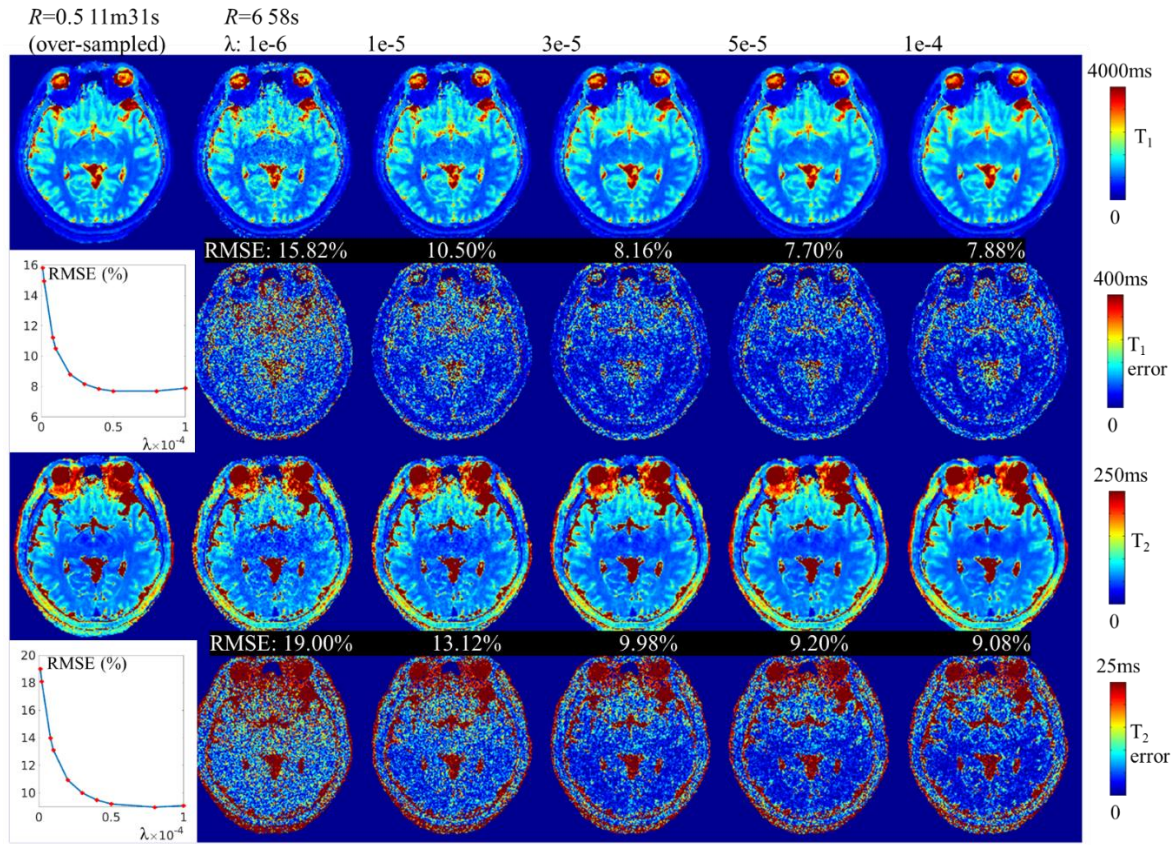

Supporting Figure S2.

$T_1$ and $T_2$ maps using the proposed TGAS-SPI-MRF at $R=6$ with LLR regularization parameter $\lambda$ from $10^{-6}$ to $10^{-4}$ using in-vivo data.

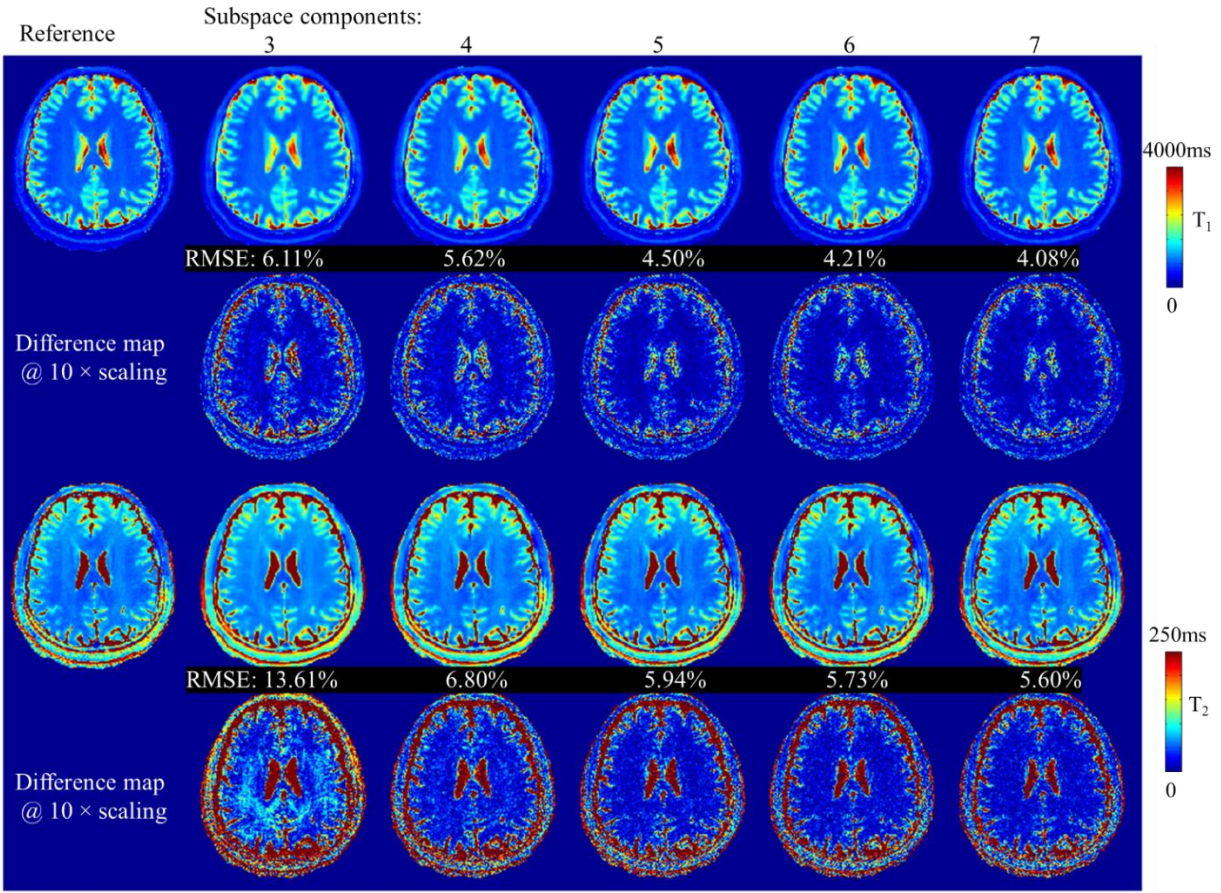

Supporting Figure S3.

$T_1$ and $T_2$ maps using the proposed TGAS-SPI-MRF at $R=3$ with different numbers of subspace components from 3 to 7 using simulated data.

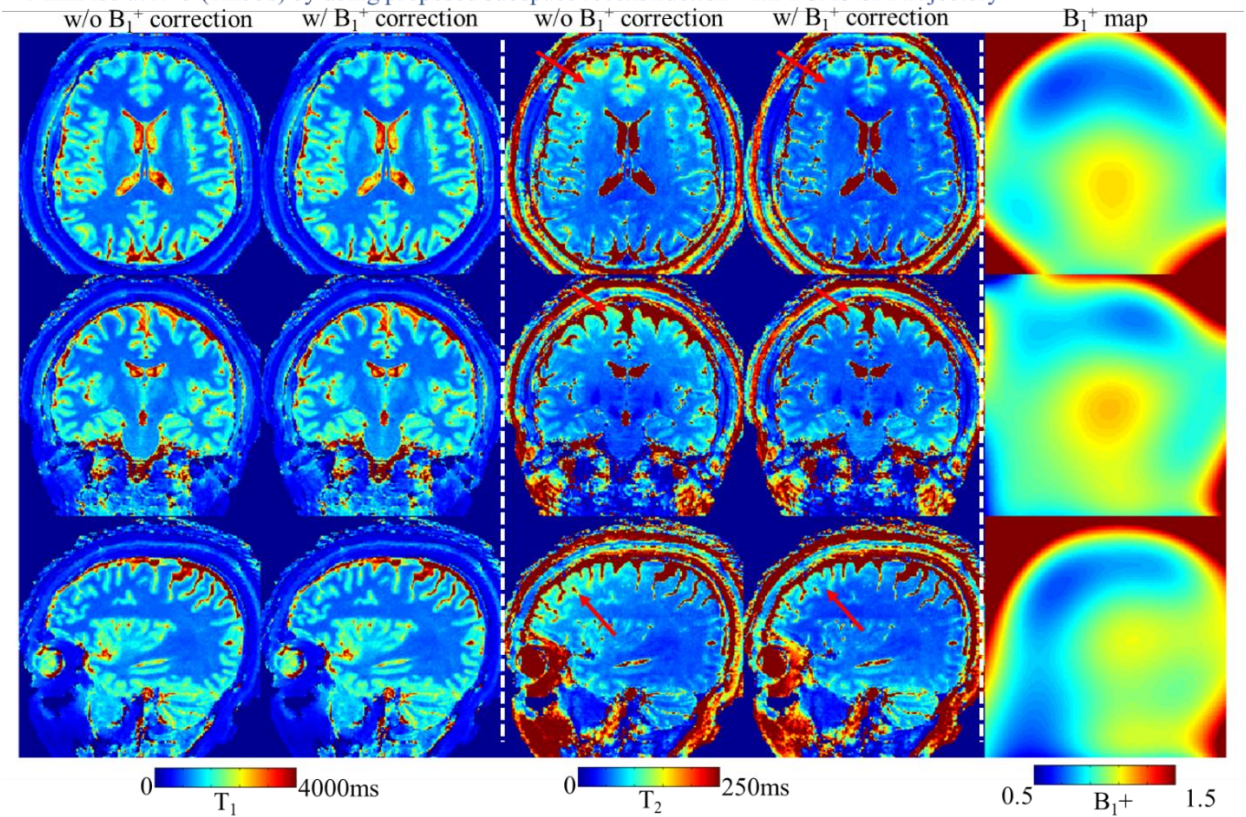

Supporting Figure S4.

$T_1$ & $T_2$ maps using the proposed TGAS-SPI-MRF with and without $B_1^+$ correction as well as corresponding $B_1^+$ maps (right-most column). While $T_1$ maps keep consistent, the $B_1^+$ correction help to improve the uniformity of $T_2$ maps in specific regions (red arrows).